\documentstyle[12pt]{article}

\begin{document}

\def\singlespace {\smallskipamount=3.75pt plus1pt minus1pt
                  \medskipamount=7.5pt plus2pt minus2pt
                  \bigskipamount=15pt plus4pt minus4pt
                  \normalbaselineskip=15pt plus0pt minus0pt
                  \normallineskip=1pt
                  \normallineskiplimit=0pt
                  \jot=3.75pt
                  {\def\smallskip {\vskip\smallskipamount}}
                  {\def\medskip   {\vskip\medskipamount}}
                  {\def\bigskip   {\vskip\bigskipamount}}
                  {\setbox\strutbox=\hbox{\vrule
                    height10.5pt depth4.5pt width 0pt}}
                  \parskip 7.5pt
                  \normalbaselines}

\begin{center}
{\bf {\large A note on the thermodynamics of gravitational radiation}}

\smallskip

\bigskip
\bigskip

{\bf {\large T. Padmanabhan$^{a}$\footnote{e-mail address: nabhan@iucaa.ernet.in} and
T. P. Singh$^{b}$\footnote{e-mail address: tpsingh@tifr.res.in} 
}}

\bigskip
\bigskip
{\it $^{a}$Inter-University Center for Astronomy and Astrophysics}\\
{\it Post Bag 4, Ganeshkhind, Pune 411007, India}\\
\medskip
{\it $^{b}$Tata Institute of Fundamental Research,}\\
{\it Homi Bhabha Road, Mumbai 400 005, India.}

\end{center}

\bigskip
\bigskip

\begin{abstract}
\noindent It is shown that linearized gravitational radiation confined in a cavity can
achieve thermal equilibrium if the mean density of the radiation and the size of the cavity satisfy certain constraints.
\end{abstract}

Is it possible, in principle, to confine gravitational radiation in a box,
for a time long enough for it to achieve thermal equilibrium? This issue
has been investigated before by various authors \cite{one, two, three}, not all of whom have arrived
at the same conclusion. The purpose of this note is to pursue a particular
line of thought which does not seem to have been looked at in earlier work.
We argue that it is possible to thermalize gravitational radiation, starting
from a non-equilibrium configuration, so long as certain constraints (to be
discussed below) are satisfied.

Let us begin by summarizing the arguments given by Smolin \cite{one}. He showed, by
considering various mechanisms of absorption of radiation by matter, that
no realistic material is an efficient absorber of gravitational radiation,
except over a narrow band of frequencies. These mechanisms include absorption
of classical gravitational radiation by classical matter, ionization of atoms 
by gravitons, and phonon excitation. He concluded that a state of thermal
equilibrium between radiation and matter cannot be reached in a finite time.

Garfinkle and Wald \cite{two} presented a counter-example to the result by using a shell
of charged matter balanced just outside its Schwarzschild radius. However
Dell \cite{three} later argued that such a shell is unstable in the presence of 
electromagnetic radiation. 

Whatever the nature of the confining cavity, the absorption coefficient $f$
will always be smaller than one. If the box is made thick enough to raise $f$
to one, it undergoes gravitational collapse.

However, it is not absolutely essential, for the purpose of attaining 
equilibrium, to have $f$ close to unity. For any cavity, it is possible to
define a leakage time-scale $t_{L}$ over which, say, a fraction $1/e$ of the
initial radiation escapes. Also, for an initial configuration 
$\rho({\nu})$ - the energy-density of gravitational radiation in the 
cavity - an equilibrium time-scale $t_{E}$ can be defined, over which the 
radiation may achieve equilibrium. What needs to be settled is whether
$t_{E}$ can be smaller than $t_{L}$, for a general initial configuration
$\rho({\nu})$. We argue below that this is possible, and $t_{E} < t_{L}$
need not imply gravitational collapse of the cavity or of the radiation inside
it. Moreover, we have no need to concern ourselves with any specific 
mechanism for confining radiation.  

Our argument is analogous to the standard discussion of Brownian motion of a molecule in a fluid, where one shows that as a consequence of the random collisions and viscous drag, the molecule attains a Maxwellian velocity distribution. A variant of this argument was used by Einstein \cite{four} to derive the Planckian distribution for electromagnetic radiation in equilibrium in a cavity. The argument went as follows. He assumed that the radiation was in equilibrium with a molecule (say, with two energy levels). The molecule absorbs and re-emits at a characteristic frequency $\nu_{0}$. The momentum transfers to the molecule take place through random collisions with the quanta of radiation, and through the systematic drag force it experiences because it sees a Doppler-shifted radiation while in motion. Einstein showed that if the molecule has a Maxwellian velocity distribution, then the exchange of momenta with radiation preserves this distribution only if the energy density of the radiation obeys the Planck law.

In his paper, Einstein concerned himself with the equilibrium situation. The discussion of momentum transfers to the molecule assumes greater significance in a study of ${\it approach}$ to equilibrium. Now, there is a systematic evolution of the r.m.s. velocity of the molecule, which may be approximately described by the Langevin equation
\begin{equation}
M\frac{d{\bf v}}{dt} = - \alpha {\bf v} + {\bf \beta} (t),
\end{equation} 
$\alpha$ being the drag coefficient and $\beta$ the fluctuating force
($\alpha$ is positive). Standard discussions \cite{five} of Brownian motion show that the molecule attains a Maxwellian distribution over a time-scale
\begin{equation}
t_{E} \approx \frac{M}{\alpha}.
\label{lan}
\end{equation}

In particular, the `fluid' pushing the molecule around could be electromagnetic radiation, and (\ref{lan}) will continue to hold. [The assumptions which go into using the Langevin equation as well as subtleties related to the
dependence of the drag coefficient $\alpha$ deserve discussion; this is presented in the last section since they
are unrelated to the main theme of this paper.] Once the Maxwellian distribution has been reached for the molecule, Einstein's argument can be applied to arrive at Planck's law for the radiation.

We now return to the case of linearized gravitational radiation in a cavity, interacting with a two-level atom. 

To begin with, we compare this system with the one consisting of an atom in electromagnetic radiation. The only difference between the two kinds of radiation is that the former is described by a spin-2 field, while the latter is spin one. As regards the two-level atom it is sufficient for us to assume that it has a non-zero cross section $\sigma$ for the absorption of gravitons at a frequency $\nu_{0}$ in its rest frame. The next point concerns the walls of the confining cavity. For the case of electromagnetic radiation it is safe to assume that one can construct walls with arbitrarily high absorptivity, and that the leakage time $t_{L}$ can be pushed to infinity. For the gravitational case, we assume that the cavity has an absorption coefficient $f({\nu})\equiv1-\eta({\nu})$, so that for a spherical cavity of radius $R$
\begin{equation}
\frac{d{\cal E}({\nu})/dt}{{\cal E}({\nu})} =
\frac{\eta(\nu)\rho(\nu,t)c.4\pi R^{2}}{4\pi R^{3}\rho(\nu,t)/3}
\label{rat}
\end{equation}
where ${\cal E}(\nu)$ is the energy in the cavity at frequency $\nu$.

This may be used to define a leakage time-scale 
\begin{equation}
t_{L}(\nu) \approx R/c\eta(\nu),
\label{lea}
\end{equation}
and a frequency averaged scale $\overline{t}_{L}=R/c\eta$, $\eta$ being the mean `leakage' coefficient. Clearly, any reference to the thickness or density of the absorbing material will be through $\eta(\nu)$. As expected, leakage takes more time for larger cavities.

Consider next, the motion of the atom of mass $M$, when it is pushed about by the gravitational radiation, which has a time-dependent spectrum 
$\rho(\nu,t)$. We assume that this motion can be described by the Langevin equation (\ref{lan}). This implies that the velocity of the atom will acquire a Maxwellian distribution over a time-scale $M/\alpha(\nu_{0})$, irrespective of its initial velocity. When this has happened for all possible frequencies, we can say that the radiation has also reached equilibrium. The time-scale $t_{E}$ should thus be defined as $M/\alpha_{min}$, where $\alpha_{min}$ is the minimum value of $\alpha(\nu)$, obtained by varying $\nu$. A major task is to relate $\alpha_{min}$ to an initial distribution $\rho_{0}(\nu)$ and we address this question below. Right now, we compare $t_{E}$ with the average time-scale
$\overline{t}_{L}$ defined earlier. Requiring $t_{E}<\overline{t}_{L}$ gives
\begin{equation}
R>Mc\eta/\alpha_{min}
\label{equ}
\end{equation}
as the necessary condition for equilibrium. 

We must ensure that the box is not dense enough to undergo gravitational collapse. In other words, $\eta$ is assumed to be close to one. We also need to ensure that when $R$ is chosen as in (\ref{equ}), the radiation in the cavity does not collapse to form a black hole. This requires
\begin{equation}
2G.\frac{4\pi}{3}R^{3}\frac{\overline\rho}{c^{2}} < Rc^{2}
\label{nbh}
\end{equation}
i.e., (ignoring numerical coefficients),
\begin{equation}
R^{2} < c^{4}/G\overline{\rho}
\label{nb2}
\end{equation}
$\overline{\rho}$ being the mean density of radiation. The compatibility of (\ref{equ}) and (\ref{nb2}) requires
\begin{equation}
\frac{M^{2}c^{2}\eta^{2}}{\alpha_{min}^{2}} < R^{2} < c^{4}/G\overline
{\rho}
\label{roc}
\end{equation}
which may be taken to imply that $\rho$ should satisfy
\begin{equation}
\overline{\rho} < \alpha_{min}^{2} c^{2}/GM^{2}\eta^{2}.
\label{ro2}
\end{equation}
We interpret the equations (\ref{equ})-(\ref{ro2}) as follows. If the mean density satisfies the bound (\ref{ro2}), then the radiation can be made to achieve equilibrium by choosing $R$ within the bounds given by (\ref{roc}). The situation is different from the one for electromagnetism because $\eta$ cannot be made one for electromagnetic radiation, and because we do not want 
$t_{E} < \overline{t}_{L}$ to imply gravitational collapse. Such a constraint on $\eta$ is only to be expected, because unlike for the electromagnetic case, the principle of equivalence a priori fixes the `charge to mass' ratio for gravity.

It is evident that perfect absorption $(f=1)$ is a sufficient but not a necessary condition for thermalization. This allows the construction of the above mechanism for achieving equilibrium. On the other hand, Smolin was mainly concerned with measurement of the quantum state of the gravitational field, and that certainly requires $f\approx 1$. Thus our results should be considered complimentary, rather than contradictory, to those of Smolin.
 
We end this discussion by a comment on the coefficient $\alpha_{min}$ in (\ref{ro2}). Near equilibrium, $\alpha_{min}(\nu)$ is proportional to $\nu$, and vanishes for $\nu=0$, which only means that no momentum is absorbed when the energy levels coincide. So it is reasonable to define $\alpha_{min}$ as $\alpha(\nu_{min})$ where $\nu_{min}$ is the smallest frequency for which we are interested in studying the approach to equilibrium. Clearly, this issue arises for electromagnetic radiation also, and is not peculiar to gravity.

For the equilibrium case, $\alpha({\nu})$ is simple to calculate; but for an arbitrary configuration an exact calculation is not possible. A very crude estimate for $\alpha_{min}$ may be obtained as follows. Consider an atom which is moving with a speed $V$ in an isotropic radiation with mean density $\overline{\rho}$. The average momentum transferred to the atom per unit time will be of the order 
\begin{equation}
\frac{\overline{\rho}}{c}.\frac{V}{c}.c\sigma=\overline{\rho}\sigma.\frac{V}{c}
\label{mom}
\end{equation}
$\sigma$ being the mean cross-section of the atoms to absorption of gravitons. This gives $\alpha_{min}\sim \overline{\rho}\sigma/c$, and from (\ref{ro2}) it follows that 
\begin{equation}
\overline{\rho}> GM^{2}\eta^{2}/\sigma^{2}.
\label{roa}
\end{equation}

Numerical estimates of the allowed ranges for $R$ and $\overline{\rho}$ will depend on $\sigma$, and on the nature of the confining box (through $\eta$). In principle, however, gravitational radiation with $\overline{\rho}$ exceeding the bound in (\ref{roa}) will reach equilibrium if placed in a suitable box.

Having concluded the central part of our discussion, we would like to comment on several aspects of the analysis which deserve careful understanding. 

We begin with the key technique used in the analysis, viz. the Langevin equation and its validity. Langevin equation arises in many situations dealing with stochastic forces (see e.g.,\cite{six} for a discussion of its validity and comparison with Fokker-Planck equation) and is a good assumption if the scattering process is uncorrelated and homogeneous. In this particular case we assume that the system is interacting with a bath of gravitons which can be described classically as a stochastic background of gravitational waves. While one could think of more complicated scenarios, the description based on Langevin equation is adequate for our purpose. Of course, Langevin, equation is never ``fundamental'' and is more like a phenomenological equation such as, say, Hooke's law. But since we are discussing a specific situation well within the domain of its validity, our results are trustworthy.

The second, and somewhat more subtle issue is whether $\alpha$ depends on $M$ and if so, how. This is interesting because, if $\alpha$ is proportional to $M$ with, say $\alpha=M\alpha_0(\nu)$ where $\alpha_0(\nu)$ is independent of $M$,
then some of our bounds (which depend on $\alpha/M$) will be independent of $M$. In fact, one may naively imagine that principle of equivalence will require $\alpha\propto M$ making final results independent of $M$. In reality, however,
$\alpha\propto M^2$ and there are different ways of seeing this. The first, and the simplest, is to note that in the electromagnetic case $\alpha\propto q^2$ (where $q$ is the charge) and {\it not}  $\alpha\propto q$; this is obvious since Langevin equation will treat both positive and negative charged particles at the same footing and the sign of $\alpha$ cannot depend on the sign of $q$. By analogy, the $\alpha$ for gravity should scale as $(GM)^2$. The second argument is based on the result in equation (\ref{mom}) where we showed that $\alpha\propto\sigma$. Now the lowest order scattering of gravitational waves by matter consists of matter oscillating under the infulence of the incident gravitational wave and radiating out gravitational waves due to its motion. This is completely analogous to the Thomson scattering of
electromagnetic radiation and the scattering crosssection for this process can be worked out in a straight forward
manner (Padmanabhan, 1978; unpublished). The result is as expected with $\sigma\propto (GM/c^2)^2;$ i.e, the Schwarzschild radius acts as the analogue of classical electron radius. Once again, we find that $\alpha\propto\sigma\propto M^2$.
Finally, note that the averaging of the stochastic force acting on the particle will involve the two point correlation function of the force at the lowest order. That is why $\alpha$ will scale in proportion with the power spectrum [energy density] of the stochastic background (quadratic in the gravitational wave amplitude) and --- by the same token ---
will scale as $M^2$.
This imples that the time taken to reach
equilibrium depends on the mass $M$ of the atom. One should not use principle of equivalence naively to argue otherwise; principle of equivalence has no direct role to play in this issue.
This is the same result as
in the treatment of Brownian motion via the Langevin equation. As a result,
the various bounds obtained above depend on $M$. 
Of course, the mass dependence of $\alpha$ is not germane to the main issue we are addressing.
Our main concern here
is to demonstrate that equilibrium can in principle be reached, even though it
implies that the bounds will depend on $M$.

It may appear that Eqns. (9) and (11) give conflicting bounds, the former being an upper bound and the latter a lower bound, on the density. What is happening here of course is that one does not in general know what $\alpha_{min}$ is, and (11) results from using a very rough estimate for $\alpha_{min}$.
It is also important to note that, we have tried to obtain fairly general conclusions with most plausible assumptions. In specific contexts, one may be able to strengthen the bounds or supercede them with other considerations but these will not be generic. For example, we have treated gravitational waves as a small perturbation on a flat background. While
the gravity waves being a perturbation defines the context of the problem, one can think of the effects of background being curved. If the scale length of background curvature is comparable to the mean free path of gravitons, there will be further effects. We suspect that any such effect, which involves scattering of gravitons by the curvature will only go towards randomising the processes and will not affect our conclusions but this claim is hard to establish.

Finally, we would like to speculate on some implication our results might have
for the issue of information loss in black hole evaporation. 
It is well-known that a black hole radiates a Planck spectrum for linearized gravitons also, besides the Planck spectrum for other quantum fields of lower spin. For all practical purposes the radiated particles behave like a set of particles thermalised by some physical process. On the other hand, complete evaporation of a black hole resulting in purely thermal radiation as relic will lead to well-known information loss problems and possibly problems with unitary evolution of quantum mechanical systems.

In the context of our work one could not help speculating whether the restrictions on thermalization of gravitational radiation by "normal" systems have any implications for the radiation of thermal gravitons by black holes. If the black-hole graviton system can be treated like the `gravitational 
radiation in a box' system discussed above, and if the system does not satisfy the conditions necessary for thermalization, the escaping non-thermal gravitational radiation could carry information. 

It is not clear whether one could interpret the black hole radiation as arising out of a "thermalization" process; but if it is, then the possibility opens up that the gravitons can contain at least part of the information lost when the event horizon has formed and may lead to some interesting implications for the information loss problem. This idea, though rather
speculative at this stage, deserves to be investigated further.

\end{document}